\documentclass[sigconf,nonacm]{acmart}

\usepackage{hyperref}
\usepackage{url}
\usepackage{mathtools}
\usepackage{pdftexcmds}
\usepackage{stmaryrd}
\usepackage{amsthm}
\usepackage{color}
\usepackage{xspace}
\usepackage{mathpartir}

\newcommand{\vect}[1]{\overrightarrow{#1}}

\newcommand{\Null}{\mathtt{null}}

\newcommand{\imp}[0]{\Rightarrow}
\newcommand{\red}[0]{\mathrel{\leadsto}}

\newcommand{\sem}[1]{\left\llbracket {#1} \right\rrbracket}

\def\orelse{\mathbin{~|~}}

\newcommand{\setlit}[1]{\{{#1}\}}

\newcommand{\comprehension}[1]{\bigcup\setlit{#1}}

\newcommand{\isempty}{\mathtt{empty}}
\newcommand{\tuple}[1]{\langle{#1}\rangle}

\def\plwhere{\mathtt{where}}
\def\setwhere{\plwhere}
\def\plempty{\mathtt{empty}}
\def\setempty{\plempty}

\newcommand{\compop}[1]{\mathop{%
   \ooalign{%
      \hfil \raise .2ex\hbox {$\scriptscriptstyle#1$}\hfil \crcr
      {$\bigcirc$}%
   }}}

\def\kwtrue{{\mathtt{true}}}
\def\kwfalse{{\mathtt{false}}}

\def\boolty{\mathbf{bool}}
\def\intty{\mathbf{int}}

\newcommand{\none}{\mathtt{none}}
\newcommand{\some}[1]{\mathtt{some}(#1)}
\newcommand{\caseof}[4]{\mathtt{case}~#1~\mathtt{of}~(\none \Rightarrow #2 \mid \some{#3}\Rightarrow #4)}
\newcommand{\ifte}[3]{\mathtt{if}~#1~\mathtt{then}~#2~\mathtt{else}~ #3}
\newcommand{\isNull}[1]{\mathtt{isNull}(#1)}

\newcommand{\isnull}{\mathsf{isnull}}
\newcommand{\val}{\mathsf{val}}

\newcommand{\trans}[1]{\llbracket #1 \rrbracket}

\DeclareMathOperator{\FV}{FV}

\newcommand{\NRCopt}{\ensuremath{\mathrm{NRC_{opt}}}\xspace}
\newcommand{\NRCnull}{\ensuremath{\mathrm{NRC_{null}}}\xspace}
\newcommand{\NRC}{\ensuremath{\mathrm{NRC}}\xspace}

\begin{document}
\title{Comprehending nulls} \author{James Cheney} \affiliation{%
  \institution{University of Edinburgh}}
  \email{jcheney@inf.ed.ac.uk}
\author{Wilmer Ricciotti}
\affiliation{%
  \institution{University of Edinburgh}}
\email{research@wilmer-ricciotti.net}
\begin{abstract}
  The Nested Relational Calculus (\NRC) has been an influential
  high-level query language, providing power and flexibility while
  still allowing translation to standard SQL queries.  It has also
  been used as a basis for language-integrated query in programming
  languages such as F\#, Scala, and Links.  However, SQL's treatment
  of incomplete information, using nulls and three-valued logic, is
  not compatible with `standard' \NRC based on two-valued logic.
  Nulls are widely used in practice for incomplete data, but the
  question of how to accommodate SQL-style nulls and incomplete
  information in \NRC, or integrate such queries into a typed
  programming language, appears not to have been studied thoroughly.
  In this paper we consider two approaches: an \emph{explicit}
  approach in which option types are used to represent (possibly)
  nullable primitive types, and an \emph{implicit} approach in which
  types are treated as possibly-null by default.  We give translations
  relating the implicit and explicit approaches, discuss 
  handling nulls in language integration, and sketch extensions of
  normalization and conservativity results.
\end{abstract}

\maketitle

\section{Introduction}

The Nested Relational Calculus (\NRC)~\cite{buneman95tcs} is a high-level query language
providing operations for collections (sets, bags, lists, etc.),
especially \emph{comprehensions}.  In contrast to standard query
languages such as SQL, \NRC queries can be freely composed and can
construct values with nesting of record and collection types, making
it natural to use for database programming and query integration
in high-level functional languages~\cite{cooper06fmco,cheney13icfp,Syme06,quill}.  Despite this added
flexibility, \NRC queries are no more expressive than flat relational
queries when transforming flat inputs to flat outputs~\cite{wong:conservativity}.  This property,
called \emph{conservativity}, is the basis for rewriting
algorithms that map \NRC queries over flat data to SQL queries.

From the early years of the development of the relational data model
and associated query languages, the importance of supporting
incomplete information has been clearly recognized. Codd~\cite{Codd79} made an early
proposal allowing field values to be ``null'', or absent/missing, 
extending primitive operations on these values to propagate
nulls, and extending predicates to have three-valued semantics with a
third truth value, ``unknown''. 
Despite criticism~\cite{Grant08}, this approach is standard and widely used in
SQL, although these features are also easily misunderstood and result
in counterintuitive behavior that can lead to subtle bugs~\cite{guagliardo17vldb}.  
Nevertheless, almost all real databases and applications involve
nulls, so it is important for language-integrated query
mechanisms to support them.  

Most presentations of \NRC and related languages eschew nulls: base
types include integers, booleans, strings, etc. as understood in most
typed programming languages, in which there is no special null value
indicating an absent piece of data.  This makes \NRC a good fit for
integrating database queries into an ambient typed language, but a
poor fit for interfacing with actual incomplete data.  Moreover, while
SQL's approach to nulls is imperfect, a language-integrated query
system should still be able deal with them.

In this short paper, we investigate the design issues
that arise when we add null values to \NRC, highlight technical issues
whose solutions are straightforward or already known, and outline open
questions.  In particular we consider the following issues:
\begin{enumerate}
\item Should nulls be treated implicitly (like in SQL) or explicitly
  (like option values in functional languages)?
  \item Should nulls be available at any type, or just at base types?
\item Do classical results needed for translating \NRC queries to SQL
  continue to hold in the presence of nulls?
\end{enumerate}

\paragraph{Design considerations}\label{sec:design}

Our goal is to reconcile the implicit treatment of nulls in a typical
database query language (e.g. SQL) with a typed, functional host
language that lacks nulls.  We first give a toy example and discuss how it is handled
currently in three settings: Links~\cite{cooper06fmco}, Scala's Quill
library~\cite{quill}, and in LINQ in F\#~\cite{Syme06,cheney13icfp}.

Suppose we have a table containing diseases, each with identifier (integer),
name (string),
and type (integer).  The identifier and name are required
(i.e. non-nullable) but the
type is optional and nullable (some new diseases might not yet have a
known type).  To produce a web page showing
information related to diseases with a given name, we would
execute a query such as
\begin{verbatim}
SELECT * FROM diseases WHERE name = 'covid-19'
\end{verbatim}

In Links, until recently, attempting
to execute queries that
attempted to read NULLs from the type field would simply fail, because
the NULL value was not expected by the code that processes query
results.  Currently, Links allows to set a single global default value
to use in place of NULL for integer fields.

In F\#, in contrast, nullable fields in database tables or query
results are given a different type: \verb|Nullable<T>|.  A value of
type \verb|T| can be implicitly coerced to \verb|Nullable<T>|, and
this type also includes a null value.  Whether a \verb|Nullable<T>| is
null or not can be tested by checking the Boolean field
\verb|HasValue|, and if present the value can be extracted from the
\verb|Value| field.  Requesting the value of a null yields an
exception.  Primitive operators such as addition and equality
(\verb|+|, \verb|=|) are
lifted to nullable versions (\verb|?+|, \verb|?=|) that propagate nulls like
SQL does: if any input is null then the result is null.

In Quill, nullable fields are given option types, and Scala
overloading and convenient operations on option types can be
used to make it easier to write queries involving such optional data.

Obviously, the Links solution is little better than a hack: if we
wanted to deal with nulls of other base types, we would have to
provide a default value, and it isn't clear that using a single global
default in place of null values of each type is sensible.  On the
other hand, the F\# and Quill approaches appear to work reasonably
well in practice, but rely on implicit coercions and exceptions, and
still require programmers to be conscious of which fields are nullable.

If we look beyond the simple scenario above in which we are just
retrieving data (possibly including NULLs) from the database, the
situation becomes a bit more complicated.  In SQL, as mentioned above,
most primitive operations are defined so that the result is null if
any input is null; some operations such as logical connectives and
null tests depart from this pattern.  Null boolean values, also called
unknowns, provide a third truth value, resulting in behavior that can
be counterintuitive.  Moreover, it is not clear that query rewriting
laws that are valid in standard two-valued logic still hold, calling
into question whether the rewriting strategy used in Links to
normalize and generate SQL from \NRC queries is still viable. We
should also note that neither F\#'s handling of nulls via nullable
types nor Quill's treatment using option types is supported by
any formal analysis like that for basic language-integrated
query~\cite{cheney13icfp}, so it is unclear what formal
guarantees these approaches have.

A final consideration, which is not strictly necessary to deal with
the problem of incomplete data in SQL, but seems natural to consider
in a nested relational setting, is whether null values ought to be
considered only for base types (integers, strings etc.) or for the composite
\NRC types including records and collection types.  The latter approach
seems more uniform and more in the spirit of \NRC, but leads
immediately to the question whether allowing nulls at composite types
increases expressiveness, or whether the classical results on
conservativity still hold.  

Summing up, we would like to reconcile database
queries involving nulls with typed host languages so that:
\begin{enumerate}
\item Null values are available at all types and query results
  including nulls can be translated to host language values.
\item Query expressions can be written as in SQL:
  e.g. primitive operations apply uniformly to nullable and nonnullable fields
\item Query expressions admit normalization rules similar to those for
  plain \NRC, enabling translation to SQL.
\end{enumerate}
Moreover, we would like to accomplish these goals in a way that makes
programming as easy as possible in common cases, and that avoids
reliance on advanced programming language features as much as possible.
We note again that none of the approaches we are aware of in Links,
F\# or Quill 
satisfy all three criteria.  

\section{Background}

We will employ the following syntax for \NRC:
\[\begin{array}{rrcl}
  \text{\textbf{Types}} &\sigma, \tau & ::= & b \orelse \tuple{\vect{\ell : \sigma}} \orelse \setlit{\sigma} 
   
  \\
  \text{\textbf{Terms}} & M, N & ::= & x \orelse c \orelse f(\vect{M}) \orelse \tuple{\vect{\ell = M}} \orelse M.\ell  \\
  & &\orelse & \emptyset \orelse \setlit{M} \orelse M \cup N
                  \orelse \bigcup\setlit{M | x \gets N} 
\\ &&\orelse& \setempty(M) \mid \ifte{M}{N_1}{N_2}
\end{array}\]
The base types $b$ include integers, strings, booleans,
floating-point numbers, dates, etc.  Constants $c$ and primitive
operations $f$ operate on base types, and include Boolean constants and logical connectives $\kwtrue,\kwfalse,\wedge,\vee,\neg$.
Record types are written
$\tuple{\vect{\ell : \sigma}}$ with records constructed as
$\tuple{\vect{\ell = M}} $ and field projection written $ M.\ell $.  
 We consider a single set collection
type written $\setlit{\sigma}$.  The
expressions involving collections include the empty collection
$\emptyset$, singleton $\setlit{M}$, union $M \cup N$, and
comprehension $\bigcup\setlit{M |x \gets N}$ where $M$ is evaluated
repeatedly with $x$ bound to elements of
$N$ and the resulting collections are unioned.  Finally, the conditional $\ifte{M}{N_1}{N_2}$ has
the standard behavior.  

We write $M ~\setwhere~ N$ to abbreviate $\ifte{N}{M}{\emptyset}$,
i.e. return $M$ if $N$ holds, otherwise
$\emptyset$.  A general comprehension  (where
$M$ may have any type)
$\setlit{M | x_1 \gets N_1,\ldots x_k\gets N_k ~\setwhere~ P}$, is syntactic sugar for
$\bigcup\setlit{\cdots \bigcup\setlit{\setlit{M} ~\setwhere~ P | x_k
    \gets N_k}\cdots | x_1 \gets N_1}$.  Such comprehensions
correspond to conjunctive SQL queries.

The (largely standard) type system and common rewriting rules for evaluating and translating
queries in this variant of \NRC are included in 
the appendix.

\section{Explicit nulls}

We extend the core \NRC with explicit nulls, calling this calculus
\NRCopt, as follows.  
\[\begin{array}{rrcl}
  \textbf{Types} & \sigma, \tau & ::= & \cdots \mid \tau?
  \\
  \textbf{Terms} & M, N & ::= & \cdots \mid \none \mid \some{M} \\
&&\mid& \caseof{M}{N_1}{x}{N_2} 
\end{array}\] 
We introduce a
new type $\tau?$ (pronounced ``$\tau$ option'') whose values are
$\none$ and $\some{V}$ where $V$ is of type $\tau$.  The elimination
form for $\tau?$ is the case construct $\caseof{M}{N_1}{x}{N_2}$ which
inspects $M$, and returns $N_1$ if $M$ is $\none$ and $N_2[V/x]$ if $M$
is $\some{V}$.  Intuitively, optional values correspond to nullable
values in SQL.  Thus, given a table with some nullable fields, these
fields can be represented using option types, whereas non-nullable
fields are represented using an ordinary type.

The semantics of option types and expressions is 
standard:
\[\begin{array}{rcl}
\caseof{\none}{N_1}{x}{N_2} &\red& N_1\\
\caseof{\some{M}}{N_1}{x}{N_2} &\red& N_2[M/x]
\end{array}\]
Thus, \NRCopt essentially models the Quill approach, but the advanced
features of Scala that make it more palatable are absent.

\section{Implicit nulls}

The explicit calculus \NRCopt provides a correct, and implementable,
strategy for handling incomplete information: we simply map nullable
types in database tables to option types, and require the query to
perform any case analysis.  However, making nulls explicit using option types is not cost-free: in the
unfortunately all-too-common case where the database schema does not
specify fields as nonnull (even if they are in practice never null),
the programmer is forced to program defensively by handling both the
$\none$ and $\some{}$ cases for each field used by the query.  This is
especially painful when performing primitive operations on multiple
nullable values: for example to simulate SQL's behavior when adding
two integers that might be null, we need to perform case analysis on
the first one, then a sub-case analysis on the second one.

In this section we consider an
alternative approach, $\NRCnull$, in which all base types are treated as including an extra value
$\Null$.  The semantics of primitive operations is augmented to handle
null value inputs; in most cases, if any input value is null then the
result is null.  The exceptions are the logical connectives, which are
instead equipped with three-valued semantics (e.g. $\kwfalse \wedge \Null
= \kwfalse$), and  operations such as $\isNull{M}$ 
that inspect a possibly-null primitive value and
test whether it is null.

The syntax of \NRCnull is \NRC extended with a null constant and with
a nullness test, as follows.  We assume the presence of primitive operations
including at least the logical connectives $\wedge,\vee,\neg$.
\[\begin{array}{rrcl}
  \textbf{Terms} & M, N & ::= & \cdots \mid \Null \mid \isNull{M}
\end{array}\] 
We do not allow nulls at record or collection types.  For collection
types in particular, the expected behavior of nulls is unclear.
The semantics of logical connectives is three-valued, as in SQL.  The
semantics of other primitive operations is strict: if any argument is
null then the result is too, otherwise the primitive operation is
performed on the non-null inputs.  Finally, if the Boolean in a
$\plwhere$ statement is null, then the statement evaluates to an empty collection (similarly to $\kwfalse$ and contrary to $\kwtrue$). This behavior can be specified by
adding the following rewriting rules to the standard NRC ones:
\begin{eqnarray*}
\isNull{\Null} &\red& \kwtrue \qquad 
\isNull{c} \red \kwfalse\\
f(\ldots\Null\ldots) &\red& \Null \qquad 
M~\plwhere~\Null \red \emptyset
\end{eqnarray*}

\section{Translations}

The implicit and explicit approaches have complementary advantages.
\NRCopt is essentially a special case of the nested
relational calculus with binary sum types.
However, if many fields are
nullable,  writing queries in \NRCopt is excruciating.
On the other hand, \NRCnull
seems easier to relate to plain SQL queries, and
writing queries that operate over possibly-null values is more
straightforward (albeit with the same pitfalls as SQL),
but normalization results for \NRC with implicit nulls do not follow immediately
from prior work.
We consider translations in each direction.

\paragraph{From \NRCopt to \NRCnull}
The main issue arising in this translation is the fact that option types
can be nested inside other type constructors, including options: for example $(\intty?\times \boolty)?$ represents an
optional pair the first element of which is also optional.  To deal
with this generality, we translate options and cases as follows:
\begin{eqnarray*}
\trans{\tau?} &=& \tuple{\isnull:\boolty,\val:\trans{\tau}}\\
\trans{\none} &=& \tuple{\isnull=\kwtrue, \val=d_{\trans{\tau}}}\\
\trans{\some{M}} &=&
                     \tuple{\isnull=\kwfalse,\val=\trans{M}}\\
\left\llbracket\begin{array}{l}\caseof{M\\}{N_1\\}{x}{N_2} \end{array}\right\rrbracket&=&
                                                                   \begin{array}{l}\ifte{M.\isnull\\
}{\trans{N_1}\\
}{\trans{N_2}[\trans{M}.\val/x]}
\end{array}
\end{eqnarray*}
where $d_\tau$ is a default value of type $\tau$.  Note that nulls,  $\isNull{-}$ and other null-sensitive primitive
operations are not needed to handle options, assuming that there are
constants of each base type in \NRCopt: this translation actually maps
$\NRCopt$ to plain \NRC.

\paragraph{From \NRCnull to \NRCopt}
Types are translated as follows:
\[
\trans{b} =b? \qquad 
\trans{\tuple{\vect{\ell:\tau}}} =
                                      \tuple{\vect{\ell:\trans{\tau}}} \qquad
\trans{\setlit{\tau}} = \setlit{\trans{\tau}}
\]
The most interesting cases of the term translation are:
\begin{eqnarray*}
\trans{c} &=& \some{c}\\
\trans{f(M_1,\ldots,M_n)} &=& f^*(\trans{M_1},\ldots,
                              \trans{M_n})\\
\trans{\Null} &=& \none\\
  \trans{\isNull{M}} &=& \trans{M} = \none\\
\trans{\ifte{M}{N_1}{N_2}} &=&
                               \ifte{\mathsf{isTrue}(\trans{M})}{\trans{N_1}}{\trans{N_2}}
\end{eqnarray*}
Here $f^*$ is the primitive operation $f$ lifted to apply to options,
i.e. $f^*(\some{v_1},\ldots,\some{v_n}) = \some{f(v_1,\ldots,v_n)}$ and
otherwise $f^*(\ldots \none \ldots) = \none$.  
These operations are definable in \NRCopt, as are
the other null-sensitive operations such as equality and logical connectives.
Conditionals must be translated so that the then-branch is
executed only if the test is true, and the else-branch if the test is
false or null.  To ensure this we use the auxiliary operation
$\mathsf{isTrue}(x) = \caseof{x}{\kwfalse}{y}{y}$.

\section{Handling nulls in query results}

The translations above establish that \NRCopt and \NRCnull are equally
expressive (and equally expressive to \NRC provided all base types
have default values).  In principle one could allow programmers to
write queries in \NRCnull, generate and evaluate the corresponding SQL
queries, and translate the results at the end to
host language values involving options.  How can we make it easy to
work with these results in a host language where field types do not have nulls?

\paragraph{Nullable type tracking} This idea is a slightly
strengthened form of F\#'s approach.  The type system could be
extended to track nullability information in queries, and using this
information try to minimize the amount of optional tagging that must
be added.  In particular, this approach could cope with the overloaded
behavior of primitive operations on nulls, by giving them types that
indicate that the result may be null only if one of the inputs may be
null; if all inputs are nonnull then so is the result.  This approach
could be encoded using a sufficiently rich type system, e.g. dependent
types or Haskell's type families. However, if schemas lack accurate information about
nullability, any benefits may be limited.

\paragraph{Null handlers} This idea is loosely inspired by the common
language feature of exception handling, and by Quill's pragmatic
approach to dealing with optional values inside queries.  Given a
query returning flat records in \NRCnull, we could consider a small
domain-specific language of \emph{null handlers} that specify how to
map the result to an \NRCopt value.  A null handler is a record of
instructions defining what to do with each possibly-null field:
\begin{enumerate}
\item $\mathsf{optional}$: return an option value
\item $\mathsf{required}$: skip this record if this field is null
\item $\mathsf{default}~v$: return default value $v$ if null
\end{enumerate}
Syntactic sugar for declaring multiple fields optional or required may
also be useful.  Of course, it is possible to provide any other
desired behavior by returning all nullable results as optional values.
If nulls are tracked by the type system, then fields that are
certainly nonnull do not need to be mentioned.

For example, the disease table query from Section~\ref{sec:design}
could have (among others) two handlers:
\[\begin{array}{c}
    \tuple{\mathsf{id}:\mathsf{required},
  \mathsf{name}:\mathsf{required}, \mathsf{type}:\mathsf{default}~-1}\\
\tuple{\mathsf{id}:\mathsf{required},
    \mathsf{name}:\mathsf{required}, \mathsf{type}:\mathsf{required}}
  \end{array}\]
The first one will use $-1$, an invalid type value, if a type field is
null, while the second will skip any records that contain null type
fields.  Nulls in the id and name fields could also lead to records being
dropped, but should not occur according to the schema.  These handlers
can be desugared to case analyses using $\isNull$ (on
the database side) or $\mathsf{case}$ (in the host language).  By
desugaring to database-side case analyses, the handling can be
performed in the database, possibly saving effort.

\section{Related and future work}

Though nulls and incomplete information have been studied extensively
for traditional query languages over flat data (see Libkin~\cite{libkin14pods} for a
recent overview), these features
appear to have attracted limited interest in the setting of
nested relational calculus or complex object query languages.  The only
work in this direction we know of is from the early years of
`non-first-normal-form' databases~\cite{levene93tods,roth89ai}.  Roth et
al.~\cite{roth89ai} studied nested relations with several variants of
nulls, including no-information, does-not-exist, and unknown, while
Levene and Loizou~\cite{levene93tods} considered only a single
`no-information' null, however
neither of these approaches corresponds exactly to the treatment of nulls in SQL, as
formalized recently by Guagliardo and Libkin~\cite{guagliardo17vldb}.

Sum types (of which $\tau?$ is
a special case) were studied in an \NRC setting by Wong~\cite{wong:conservativity}.  Wong showed normalization and conservative extension
properties hold in the presence of sums and later Giorgidze et al.~\cite{giorgidze13ddfp} showed that
nonrecursive algebraic data types (i.e. n-ary labeled sums) can be
implemented in \NRC by mapping such datatypes to nested collections.
However, for the purposes of normalizing queries and generating SQL,
the latter approach has the disadvantage that query results would use
nested collections to represent options, requiring a further
flattening or shredding step possibly resulting in executing several
SQL queries~\cite{ulrich19phd,cheney14sigmod}, which is not needed in
our translation.  General sum types can also be simulated using
options, e.g. by representing $\tau + \sigma$ as
$\tuple{L:\tau?,R:\sigma?}$.  Implementing sum types using nulls is
possible future work.

In this paper we have focused on nulls in a conventional \NRC with a
single collection type, e.g. homogeneous sets or multisets.  In SQL,
which contains operators with both set and multiset semantics, as well
as grouping and aggregation, nulls interact with several other
features, such as multiset difference and aggregation, often in counterintuitive ways~\cite{Benzaken19,guagliardo17vldb}.  
Our focus has been on semantics of \NRC queries in the presence of
nulls. We conjecture that normalization and conservativity results
hold for \NRCnull and \NRCopt facilitating their translation to flat
SQL queries.  We are also interested in generalizing our treatment of
nulls to queries over heterogeneous (set/bag)
collections~\cite{ricciotti19dbpl}, higher-order
functions~\cite{Cooper09,ricciotti20fscd}, grouping and
aggregation~\cite{okura20flops}, and to shredding queries that produce
nested results into multiple SQL
queries~\cite{cheney14sigmod,ricciotti21esop} and in extending
\NRCnull to allow nulls at record and collection types.  Such extensions seem possible but not necessarily
straightforward.  For example, should a union  of a null collection
with another be null, or should the result retain partial
knowledge about the known elements?

\section{Conclusions}

Incomplete information is needed in most real database situations.
While incomplete information has been studied extensively  both in
theory (e.g. certain answer semantics~\cite{libkin14pods}) and practice (e.g SQL's
pragmatic, but complex treatment of nulls and three-valued logic~\cite{guagliardo17vldb}), almost all such
work has focused on conventional, flat relational data and queries, not
nested relations.  This gap in the literature is particularly noticeable where clean query
languages such as \NRC are used to embed SQL queries safely into an ambient
typed programming language, as in Links, F\#, or Quill.  
In this short
paper, we have outlined the main issues and design considerations we
think are important for a satisfactory solution to this problem.  We
have also outlined some initial technical steps towards a solution.

\begin{acks}
  This work was supported by ERC Consolidator Grant Skye (grant number 
  ERC 682315), and by an ISCF Metrology Fellowship grant provided by 
  the UK government’s Department for Business, Energy and 
  Industrial Strategy (BEIS).
\end{acks}
\bibliography{paper}
\bibliographystyle{abbrv}

\newpage
\appendix
\section{Typing rules}

\subsection{Rules for \NRC}

\begin{mathpar}
\inferrule{x : \tau \in \Gamma}{\Gamma \vdash x : \tau}
%
%
%
\and
\inferrule{\Sigma(c) = b}
{\Gamma \vdash c : b}
\and
\inferrule{\Sigma(f) = \vect{b_n} \imp b'
\\
(\Gamma \vdash M_i : b_i)_{i = 1,\ldots,n}}
{\Gamma \vdash f(\vect{M_n}) : b'}
\and
\inferrule{(\Gamma \vdash M_i : \tau_i)_{i = 1, \ldots, n}}
{\Gamma \vdash \tuple{\vect{\ell_n = M_n}} : \tuple{\vect{\ell_n : \tau_n}}}
\and
\inferrule{\Gamma \vdash M : \tuple{\vect{\ell_n : \tau_n}}
\\
i \in \setlit{1,\ldots,n}}
{\Gamma \vdash M.\ell_i : \tau_i}
\and
\inferrule{ }{\Gamma \vdash \emptyset : \setlit{\tau}}
\and
\inferrule{\Gamma \vdash M : \tau}
{\Gamma \vdash \setlit{M} : \setlit{\tau}}
\and
\inferrule{\Gamma \vdash M : \setlit{\tau}
\\
\Gamma \vdash N : \setlit{\tau}}
{\Gamma \vdash M \cup N : \setlit{\tau}}
\and
\inferrule{\Gamma, x:\sigma \vdash M : \setlit{\tau}
\\
\Gamma \vdash N : \setlit{\sigma}}
{\Gamma \vdash \comprehension{M | x \leftarrow N} : \setlit{\tau}}
\and
\inferrule{\Gamma \vdash M : \setlit{\tau}}
{\Gamma \vdash \setempty(M) : \boolty}
\and
\inferrule{\Gamma \vdash M : \boolty
\\
\Gamma \vdash N_1 : \tau
\\
\Gamma \vdash N_2 : \tau}
{\Gamma \vdash \ifte{M}{N_1}{N_2} : \tau}
\end{mathpar}

\subsection{Additional rules for \NRCopt}
For \NRCopt the following typing rules are added to those of \NRC:
\[\inferrule{\strut}{\Gamma \vdash \none : \tau?}\qquad \inferrule{\Gamma \vdash M : \tau}{\Gamma \vdash \some{M} :
  \tau?}\]
\[\inferrule{\Gamma \vdash M : \tau? \\
\Gamma \vdash N_1 : \sigma\\
\Gamma,x:\tau \vdash N_2 : \sigma}{
\Gamma \vdash \caseof{M}{N_1}{x}{N_2}: \sigma}
\]

\subsection{Additional rules for \NRCnull}
For \NRCnull the following typing rules are added to those of \NRC:
\[\inferrule{\strut}{\Gamma \vdash \Null : b}\qquad \inferrule{\Gamma \vdash M : b}{\Gamma \vdash \isNull{M} :
  \boolty}\]

\section{Rewrite rules}

\subsection{Common rules}

\begin{mathpar}

\tuple{\ldots, \ell = M, \ldots}.\ell \red M

\and

f(\vect{V}) \red \sem{f}(\vect{V})

\and

\comprehension{\emptyset | x \leftarrow M} \red \emptyset

\and

\comprehension{M | x \leftarrow \emptyset} \red \emptyset

\and

\comprehension{M | x \leftarrow \setlit{N}} \red M[N/x]

\and

\comprehension{M \cup N | x \leftarrow R} \red \comprehension{M | x \leftarrow R} \cup \comprehension{N | x \leftarrow R}

\and

\comprehension{M | x \leftarrow N \cup R} \red \comprehension{M | x \leftarrow N} \cup \comprehension{M | x \leftarrow R}

\and

\begin{array}{l}
\displaystyle
\comprehension{M | y \leftarrow \comprehension{R | x \leftarrow N }} 
\\
\quad \red 
\displaystyle
\comprehension{\comprehension{M | y \leftarrow R} | x \leftarrow N} \quad \mbox{(if $x \notin \FV(M)$)}
\end{array}

\and

\comprehension{M | x \leftarrow N~\plwhere~L} 
\red
\comprehension{M | x \leftarrow N}~\plwhere~L

\and 

M~\plwhere~\kwtrue \red M

\and

M~\plwhere~\kwfalse \red \emptyset

\and 

\emptyset~\plwhere~L  \red \emptyset

\and

(M \cup N)~\plwhere~L \red (M~\plwhere~L) \cup (N~\plwhere~L)

\and

\begin{array}{l}
\displaystyle
\comprehension{M | x \leftarrow N}~\plwhere~L 
\\
\quad \red 
\displaystyle
\comprehension{M~\plwhere~L | x \leftarrow N}
\quad \mbox{(if $x \notin \FV(L)$)}
\end{array}

\and

(M~\plwhere~L_1)~\plwhere~L_2 \red M~\plwhere~(L_1 \land L_2)

\and

\isempty~M \red  \isempty~(\comprehension{\tuple{} | x \gets M}) \quad \mbox{(if $M$ is not relation-typed)}

\and

\begin{array}{l}
\ifte{L}{M}{N} \red \tuple{\vect{\ell = \ifte{L}{M.\ell}{N.\ell}}} 
\\
\multicolumn{1}{r}{\mbox{(if $M,N$ have type $\tuple{\vect{\ell : \sigma}}$)}}
\end{array}

\end{mathpar}

\subsection{Additional rule for \NRC}
For \NRC, the following rewrite rule is added to the common rules:
\[
\begin{array}{l}
\ifte{L}{M}{N} \red (M~\plwhere~L) \cup (N~\plwhere~{\lnot L}) 
\\
\multicolumn{1}{r}{\mbox{(if $M,N$ have type $\setlit{\sigma}$ and $N \neq \emptyset$)}}
\end{array}
\]

\subsection{Additional rules for \NRCopt}
For \NRCopt, the following rewrite rules are added to those of \NRC:
\begin{mathpar}
\caseof{\none}{N_1}{x}{N_2} \red N_1
\and
\caseof{\some{M}}{N_1}{x}{N_2} \red N_2[M/x]
\end{mathpar}

We believe that the case analysis rules, being a special case of sum types, are well behaved and preserve the strong normalization property.

\subsection{Additional rules for \NRCnull}
For \NRCnull, the following rewrite rules are added to the common rules:
\begin{mathpar}
M~\plwhere~\Null \red \emptyset
\and
\begin{array}{l}
\ifte{L}{M}{N} 
\\
\quad \red (M~\plwhere~L) \cup (N~\plwhere~(\isNull{L} \lor {\lnot L})) 
\\
\multicolumn{1}{r}{\mbox{(if $M,N$ have type $\setlit{\sigma}$ and $N \neq \emptyset$)}}
\end{array}
\and
\isNull{\Null} \red \kwtrue
\and
\isNull{c} \red \kwfalse
\and
f(\ldots\Null\ldots) \red \Null
\and
(M~\plwhere~\Null) \red \emptyset
\end{mathpar}

Notice that the $\mathsf{if}$-splitting rule is refined to account for the case where the condition is $\Null$; this additional check preserves Girard-Tait reducibility and we thus believe the rewrite system to be strongly normalizing.

\end{document}